\documentclass[USenglish]{article}

\usepackage[utf8]{inputenc}%(only for the pdftex engine)
\usepackage[small]{dgruyter}
\usepackage{microtype}
\usepackage{adjustbox}
\usepackage{amsmath}
\usepackage{bm}
\usepackage{totcount}
\regtotcounter{table}
\regtotcounter{figure}

\begin{document}

  \articletype{Research Article}

  \author*[1]{Fabrizio Di Mari}
  \author[2]{Roberto Rocci}
  \author[3]{Silvia Rossi}
  \author[4]{Giovanna Tagliabue}
  \author[3]{Roberta De Angelis}
  \runningauthor{Fabrizio Di Mari}
  \affil[1]{Dipartimento di Economia, Statistica e Finanza,
	Università della Calabria,
	Via Pietro Bucci, 87036, Arcavacata CS, Italy, \texttt{fabrizio.dimari@unical.it}. \newline 
    Department of Medical Epidemiology and Biostatistics, Karolinska Institutet, Nobels väg 12a, 17165 Solna, Sweden.}
  \affil[2]{Dipartimento di Scienze Statistiche, Sapienza Università di Roma, Piazzale Aldo Moro 5, 00185, Rome, Italy.}
  \affil[3]{Dipartimento di Oncologia e Medicina Molecolare, Istituto Superiore di Sanità, Viale Regina Elena 299, 00161, Rome, Italy.}
  \affil[4]{Registro Tumori, Fondazione IRCCS Istituto Nazionale dei Tumori, Via Augusto Vanzetti 5, 20133, Milan, Italy.}
  \title{Flexible Estimation of the Heterogeneous Non-Parametric Component in a Relative Survival Cure Model}
  \runningtitle{Flexible Functions in a Relative Survival Cure Model}
  \subtitle{}
  \abstract{Estimating the cure fraction in a diseased population is crucial for both patients and clinicians. It offers a valuable measure for monitoring and interpreting trends in disease outcomes. When information on the cause of death is unavailable or unreliable, the Relative Survival (RS) framework is the preferred approach for estimating Net Survival (NS), which represents survival in a hypothetical scenario where the disease of interest is the only possible cause of death. In the context of cancer, NS often reaches a plateau, indicating that a portion of patients is cured, as they have the same risk of dying as a comparable group of healthy individuals with similar demographic characteristics. Classical RS cure models use logistic regression to estimate the fraction of cured patients. However, this functional form is somewhat arbitrary, and misspecifying it can severely distort the resulting cure indicators. Consequently, evaluations of the efficacy of cancer treatments at the population level could be inaccurate, leading to biased decision-making. In this paper, we address this issue by relaxing the parametric assumption and considering flexible functions of the covariates within the framework of \textit{Generalized Additive Models} (GAM) and \textit{Neural Networks}. We design an Expectation-Maximization (EM) algorithm for these cure models and conduct a simulation study to compare our proposals with the classical approach. We apply our methodology to a real-world dataset from a historical Italian cancer registry. The results demonstrate that our proposed models perform better than the classical approach and provide valuable insights into the survival outcomes of Italian colon cancer patients.}
  \communicated{...}
  \dedication{...}
  \received{...}
  \accepted{...}
  \journalname{...}
  \journalyear{...}
  \journalvolume{..}
  \journalissue{..}
  \startpage{1}
  \aop
  \DOI{...}

\maketitle

\begin{center}
\textbf{Manuscript Information} \\
\begin{tabular}{ll}
\textbf{Word count:} & 7079 \\
\textbf{Number of tables:} & \total{table} \\
\textbf{Number of figures:} & \total{figure} \\
\textbf{Supplementary material:} & No \\
\end{tabular}
\end{center}

\section{Introduction}
\label{sec:intro}
Cancer survival has dramatically improved over the last few decades and an increasing number of patients is being cured of cancer \cite{dalmaso2020}. Evidence of recovery from this severe illness has the potential to counter discrimination and social stigma. For instance, individuals cured of cancer should have the \textit{right to be forgotten} \cite{scocca}, meaning that they should have access to financial and insurance services under the same conditions as those who have never had the disease \cite{massart2018,saul2018}. 

The percentage of patients cured of cancer, as well as the time after which a patient can be considered cured, are of significant interest to public health. Indeed, through this information it can be assessed whether the introduction of a new drug or healthcare policy at the population level is effective in the effort to defeat cancer. The calculation of these indicators is possible thanks to cancer registries, which, since their establishment, have systematically collected information on all cancer patients within a specific geographical area. However, estimating these measures poses challenges. In real-world scenarios, many competing mortality causes coexist with the one under investigation \cite{tsiatis2014}, and, to provide accurate cure indicators and draw cancer-specific conclusions, it is essential to isolate the risk of death attributable only to the disease. Otherwise, we would not be able to determine whether the changes observed in patient survival are actually due to improved survival from cancer itself, or simply to patients not dying earlier from other causes. Within this context, our interest lies in estimating the survival of patients as if the disease of interest were the sole possible cause of death, an estimand known as \textit{Net Survival} (NS). In the cause-specific settings, NS can be estimated by using the cause-of-death information for each patient in the cohort, setting deaths due to cancer as the event of interest and deaths due to other causes as censoring. However, in population-based cancer registration, the cause of death is obtained from death certificates, which are often inaccurate \cite{percy1981,ashworth1991,begg2002} or even missing. 

The \textit{Relative Survival} (RS) framework \cite{ederer1961} is an alternative approach to estimate NS when cause of death information is unavailable or unreliable. In this framework two survival functions are compared: the (observed) all-cause survival in the group of diseased patients and the \textit{Expected Survival} (ES), which is the survival in a comparable group of individuals in the general population who are not under study and are considered free of the disease under investigation. By combining all-cause survival and ES, the RS framework provides an alternative way to isolate cancer-specific survival experienced by patients diagnosed with cancer. Historically, one of the most commonly used non-parametric estimator in the RS framework was the \textit{Ederer II} estimator \cite{ederer1961}. Perme et al. (2011) \cite{perme2011} showed that the Ederer II estimator is actually a biased estimator of NS, since it actually estimate another quantity, the \textit{Observable} NS, the survival function associated to the \textit{Cause-Specific Hazard} \cite{tsiatis2014}. They proposed another estimator, the \textit{Pohar-Perme} estimator, which was proved to be a consistent estimator of NS. It has been subsequently shown that using age-standardization or age-group specific estimates, the bias coming from the Ederer II estimator considered as estimator of NS is negligible, and there is also a “benefit” in precision for longer term survival \cite{lambert2015}. 

For some cancers, the above estimators plateau after a certain follow-up time, indicating that the survival of the patients becomes equivalent to that of the comparable group of individuals in the general “illness-free” population. The group still alive is therefore considered \textit{statistically cured}, although we cannot ultimately determine whether these individuals are completely free of symptoms. Cure models within the RS framework offer a model-based approach to estimate the fraction of cured patients by imposing an asymptote on the NS. Several cure models have been proposed to estimate the proportion of cancer patients that are statistically cured and they belong to two classes: the \textit{mixture} cure models, where NS is assumed to be a mixture of a flat component (the non-parametric component) and another component describing the NS of fatal patients, and the \textit{non-mixture} cure models, where NS is assumed to be the cure fraction raised to a cumulative distribution function, thus imposing an asymptote exactly equal to the cure fraction. Both family of models are able to divide the patients' population into two groups, the ones that will never experience the event of interest (cancer death), namely the cured patients, and the ones who have non-zero probability of the event of interest to occur. Within the context of RS mixture cure models, De Angelis et al. (1999) \cite{deangelis1999} used both Exponential and Weibull distributions to characterize the survival function of fatal patients. Lambert et al. (2007) \cite{lambert2007} proved how modeling both parameters of the Weibull distribution as a function of the covariates is crucial to avoid biased estimates of cure indicators. Yu et al. (2004) \cite{yu2004} explored the challenges related to identifiability in RS mixture cure models under conditions where the length of follow-up does not sufficiently support the assumption of an existing fraction of cured patients. Additionally, when this situation occurs they showed how robust the generalized gamma distribution is to model fatal cases. However, all these models share the significant limitation of not being sufficiently flexible to follow the non-parametric estimators of NS for older age groups, as they should \cite{poharperme2016}. Consequently, less restrictive parametric assumptions are needed to avoid model misspecification. Within the context of RS non-mixture cure models, Lambert et al. (2009) \cite{lambert2009} addressed this drawback by using a mixture of two Weibull distributions to characterize fatal patients. Each component represents a distinct group: those likely to die of cancer within a short period due to late diagnosis, and those with longer survival prospects resulting from earlier diagnosis. They used the same data of their previous work \cite{lambert2007} but, while they improved previous estimations modeling two groups among fatal cases, the oldest group was still not well represented. Andersson et al (2011) \cite{andersson2011} used a flexible parametric survival approach based on cubic splines with knots spread over the whole follow up time period. The cure proportion was estimated imposing the condition that the survival associated to the excess risk of dying from cancer was flat after the last placed knot. In their application they showed overall good adaptation to the non-parametric estimates even for the oldest age group, and robustness with respect to the choice of the knots. However, the cure point coincides with the last knot, and, hence, situations where the cure is not reached within the available follow-up time cannot be effectively handled with this type of models. Another way to estimate the proportion of cured patients is to use \textit{Neural Networks}. These tools have been already widely used in causal inference \cite{lechner2023}, and their usage is increasing even in survival analysis. 

Faraggi and Simon (1995) \cite{faraggi1995} were the first to model censored survival data with Neural Networks, replacing the linear predictor in the Cox model \cite{cox1972} with a simple feedforward Neural Network to capture complex nonlinearity among covariates. Many other statistical methodologies incorporating Neural Networks to model various types of longitudinal \cite{francesca2021}, survival \cite{ghosh2004, tao2022}, and high-dimensional data \cite{yeseul2024} have been proposed over the years, but Xie and Yu (2021) \cite{xie2021} were the first to use a Neural Network to model the non-parametric component of mixture cure models. Exploiting the Neural Network ability to digest complex data structures, they were able to use high-dimensional and even complex data structures, such as images or vectorized text, to model the cure probability. They also showed how, under mild conditions, the obtained estimators are consistent. 

We believe that the use of more flexible approaches in the context of cure models can help, on one hand to avoid model misspecification and, on the other hand, to maintain a solid statistical interpretation. In this paper, we propose to model the non-parametric component of the RS mixture cure model outlined in Lambert et al. (2007) \cite{lambert2007} with flexible non-parametric and semi-parametric functions of the covariates: on one hand, using one-hidden layer feedforward Neural Networks, and on the other hand \textit{Generalized Additive Models} (GAM) \cite{hastie1986} with interactions. Our primary objective is to determine whether more flexible functions of the covariates can capture potential nonlinear effects and interactions between variables in the heterogeneous cure probability. Furthermore, we derive an Expectation-Maximization (EM) algorithm \cite{dempster1977} for this type of RS cure models to find the maximum likelihood estimates of the model parameters. 

We have recently published a flexible model-based approach to divide the population of cancer patients into different subgroups based on cancer severity \cite{DiMari2025}. While in that previous study the main objective was to account for heterogeneity among uncured cancer patients, in this work we aim to provide improved estimation of cure probabilities. Another distinctive feature concerns the distributions assumed for fatal patients. In the previous study, we modeled only the scale parameter of the Weibull distributions, following the approach of De Angelis et al. (1999) \cite{deangelis1999}. In contrast, in this work we model both parameters of the Weibull distribution for the latency component, in the spirit of Lambert et al. (2007) \cite{lambert2007}. 

The remainder of this paper is organized as follows. In Section \ref{sec:meth}, we introduce the models and outline the estimation procedure. In Section \ref{sec:sim}, we compare the proposed models with the model proposed by Lambert et al. (2007) \cite{lambert2007} through an extensive simulation study under various scenarios. In Section \ref{sec:appl}, we apply our proposed models to a cohort of patients diagnosed with colon cancer from the Italian cancer registry of Varese, used in our previous work \cite{DiMari2025}. We conclude the study with a discussion.

% from here
\section{Modeling and estimation framework}
\label{sec:meth}
%\subsection{The model}
Suppose a fatal disease is affecting a population and that patients can be divided into two distinct strata: a proportion $\theta$ of cured patients with respect to the disease, having death hazard $h_{0}$, density function $f_{0}$ and survival probability $S_{0}$; and a proportion $(1-\theta)$ of patients who are bound to die of the disease, having death hazard $h_{1}$, density function $f_{1}$, and survival probability $S_{1}$. We assume the following additive model for the hazard, $h_{1}(t,\bm{x})=h_{0}(t,\bm{x}) + h_{D}(t,\bm{x})$, where $t$ and $\bm{x}$ denote the observed time to event and a vector of given covariates, respectively, and $h_{D}$ is the excess risk due to the fatal disease. The corresponding survival function $S_{1}$ is given by $S_{1}(t,\bm{x})=S_{0}(t,\bm{x})S_{D}(t,\bm{x})$, where $S_{D}$ is the survival function associated with the additional death risk $h_{D}$, which, under the independent risks assumption, represents the NS of the patients. In the RS framework, $S_{0}$ is estimated through population life tables. The use of life tables is justified by the assumption that the mortality due to the \textit{rare} disease is negligible compared to all other causes of mortality in the general population. We further assume the cure proportion to be heterogeneous in the patients population and dependent of another set of covariates $\bm{y}$ not necessarily equal to $\bm{x}$. Then, the (overall) NS mixture can be written as
\begin{equation}
\label{eq:overallNS}
    S(t,\bm{x})=\theta(\bm{y})+\{1-\theta(\bm{y})\}S_{D}(t,\bm{x}),
\end{equation}
where the NS of cured patients is identically equal to 1. 

When a linear predictor is linked with the cure probability $\theta(\cdot)$ with a logit link \cite{nelder1972}, a particular functional form is assumed. In our proposal, cure probability has been modeled as a flexible function of the covariates $\bm{y}$ using a one-hidden layer feedforward Neural Network and a GAM.

More generally, let $\theta(\cdot)$ be a function of unknown shape, and $\bm{x} \in \mathbb{R}^{p+1}$ and $\bm{y} \in \mathbb{R}^{q+1}$ our two vectors of covariates with first entry equal to one. Suppose the survival component $S_{D}$, namely the NS associated with the uncured group, has a Weibull parametric form, i.e. $S_{D}(t,\bm{x})=\text{exp}[-(\lambda(\bm{x})t)^{\beta(\bm{x})}].$
Both scale and shape parameter of the distribution are modeled as a function of covariates, as the work by Lambert et al. (2007) \cite{lambert2007} recommends, namely $\lambda(\bm{x})=\text{exp}(\bm{x}^{T}\bm{\gamma})$ and $\beta(\bm{x})=\text{exp}(\bm{x}^{T}\bm{\xi})$, where the log link ensure the positiveness of both parameters, and $\bm{\gamma} \in \mathbb{R}^{p+1}$ and $\bm{\xi} \in \mathbb{R}^{p+1}$ are unknown regression parameters. As a result, the model just described is semi-parametric. The parameters $\bm{\gamma}$, and $\bm{\xi}$ are estimated by maximum likelihood on the basis of individual data.

Let us consider $n$ patients entering a follow-up of length $\tau$. Let $t_{i}$ denote the observed time to event for the $i$-th patient, which could be either the censoring time or the time of death, as determined by the censoring indicator $\delta_{i}$ ($\delta_{i}=1$ for time to death and $\delta_{i}=0$ for time to censoring). Assuming independent and identically distributed observations, we can explicitly write the likelihood function. The contribution of the $i$-th patient to the likelihood is given by
\begin{align}
    \label{i_lik_1}
        L_{i}(\bm{\gamma},\bm{\xi} \, ; \, \bm{x}_{i}, \bm{y}_{i}, t_{i},\delta_{i}) & = \theta(\bm{y}_{i})\,S_{0}(t_{i},\bm{x}_{i})h_{0}(t_{i},\bm{x}_{i})^{\delta_{i}} + \nonumber \\ 
        & + \{1-\theta(\bm{y}_{i})\}\,S_{0}(t_{i},\bm{x}_{i})S_{D}(t_{i},\bm{x}_{i}) \cdot \nonumber \\
        & \: \cdot \{h_{0}(t_{i},\bm{x}_{i})+h_{D}(t_{i},\bm{x}_{i})\}^{\delta_{i}}, \nonumber
\end{align}
where we omitted the parameters dependence on each element of the likelihood for the sake of readability; hence, the log-likelihood for all observations is given by
\begin{equation}
\label{log-likelihood}
    \ell(\bm{\gamma},\bm{\xi} \, ; \, \bm{X}, \bm{Y}, \bm{t},\bm{\delta}) = \sum_{i=1}^{n} \text{log}\{L_{i}(\bm{\gamma},\bm{\xi} \, ; \, \bm{x}_{i}, \bm{y}_{i}, t_{i},\delta_{i})\}.
\end{equation}
%\subsection{Estimation}
%\label{sec:estimation}

We would like to maximize (\ref{log-likelihood}) to find the maximum likelihood estimates of the parameters. For this purpose, we use the EM algorithm and treat group assignment cured/uncured as unobserved latent binary variables. Let $Z_{i}$ be i.i.d. random variables, $i=1,...,n$, such that $Z_{i}=1$ if the patient $i$ is cured of the disease and $0$ otherwise; then $Z_{i} \sim$ Bernoulli$\{\theta(\bm{y}_{i})\}$. When $\{z_{i}\}_{i=1}^{n}$ are observed, the complete data log-likelihood is given by
%\begin{equation}
\begin{align}
    \ell_{c}(\bm{\gamma},\bm{\xi} \, ; \, \bm{X}, \bm{Y}, \bm{t},\bm{\delta}, \bm{z}) = 
    & \sum_{i=1}^{n} z_{i}\log\{\theta(\bm{y}_{i})\,S_{0}(t_{i},\bm{x}_{i})h_{0}(t_{i},\bm{x}_{i})^{\delta_{i}}\} \; + \nonumber \\
    & + (1-z_{i})\log[\{1-\theta(\bm{y}_{i})\}\,S_{0}(t_{i},\bm{x}_{i})S_{D}(t_{i},\bm{x}_{i}) \cdot \nonumber \\
    & \cdot\{h_{0}(t_{i},\bm{x}_{i})+h_{D}(t_{i},\bm{x}_{i})\}^{\delta_{i}}\}]. \nonumber
\end{align}
%\end{equation}
Hathaway (1986) \cite{hathaway1986} showed that the EM algorithm for mixture distributions is equivalent to apply coordinate ascent to the following function, also called \textit{fuzzy} function
\begin{equation}
\begin{split}
\label{fuzzy_func}
    \ell_{f}(\bm{u},\bm{\gamma},\bm{\xi}) = 
    & \sum_{i=1}^{n} u_{i}\log\{\theta(\bm{y}_{i})\,S_{0}(t_{i},\bm{x}_{i})h_{0}(t_{i},\bm{x}_{i})^{\delta_{i}}\} \; + \\
    & + (1-u_{i})\log[\{1-\theta(\bm{y}_{i})\}\,S_{0}(t_{i},\bm{x}_{i})S_{D}(t_{i},\bm{x}_{i}) \cdot \\
    & \cdot\{h_{0}(t_{i},\bm{x}_{i})+h_{D}(t_{i},\bm{x}_{i})\}^{\delta_{i}}] + \\
    & - u_{i}\log(u_{i}) - (1-u_{i})\log(1-u_{i}),
\end{split} 
\end{equation}
subject, in this specific case, to the constraint that all variables $u_{i}$ must lie in $[0,1]$, $\forall \; i \in \{1,..,n\}$. Coordinate ascent algorithm is based on the idea that the maximization of a multivariate function can be achieved by maximizing it along one direction at a time, i.e., solving univariate optimization problems in a loop. The first step (equivalent to the Expectation step) is to maximize (\ref{fuzzy_func}) with respect to the variables $u_{i}$, $\forall \; i \in \{1,..,n\}$, to compute the posterior probabilities, which boils down to the following update
\begin{equation*}
\begin{split}
    &\hat{u}_{i}=\mathbb{E}[Z_{i}|\bm{x}_{i},\bm{y}_{i},t_{i},\delta_{i}] = \frac{\hat{u}_{i0}}{\hat{u}_{i0}+\hat{u}_{i1}}, \\
    &\hat{u}_{i0} = \theta(\bm{y}_{i})\,S_{0}(t_{i},\bm{x}_{i})h_{0}(t_{i},\bm{x}_{i})^{\delta_{i}}, \\
    &\hat{u}_{i1} = \{1-\theta(\bm{y_{i}})\}\,S_{0}(t_{i},\bm{x}_{i})S_{D}(t_{i},\bm{x}_{i})\{h_{0}(t_{i},\bm{x}_{i})+h_{D}(t_{i},\bm{x}_{i})\}^{\delta_{i}}.
\end{split}
\end{equation*}
We now seek to optimize (\ref{fuzzy_func}) with respect to $\bm{\gamma}$, $\bm{\xi}$ and the unknown function $\theta(\cdot)$. When we evaluate the function (\ref{fuzzy_func}) on $\bm{\hat{u}}=\{\hat{u}_i\}_{i=1}^{n}$, the objective function becomes
\begin{equation}
\label{fuzzy_neural}
    \ell_{f}(\bm{\hat{u}},\bm{\gamma},\bm{\xi}) = \ell_{c}(\bm{\gamma},\bm{\xi} \, ; \, \bm{X}, \bm{Y},\bm{t},\bm{\delta}, \bm{\hat{u}}) + c,
\end{equation}
where $c$ is constant with respect to the parameters $\bm{\gamma}$ and $\bm{\xi}$, and with respect to $\theta(\cdot)$. Therefore the subsequent optimization steps (equivalent to the Maximization step) involve a coordinate ascent on the complete data log-likelihood, evaluated at the posterior probabilities $\bm{\hat{u}}$. Now, we look for a flexible function $\theta(\cdot)$ which maximizes the profile of (\ref{fuzzy_neural}), that is the sum over all observations of the negative \textit{cross-entropy} between the posterior probability distribution and the prior probability distribution, namely
\begin{equation}
\label{eq:negcrossentr}
    -H = \sum_{i=1}^{n} \hat{u}_{i}\text{log}(\theta(\bm{y_{i}})) + (1-\hat{u}_{i})\text{log}(1-\theta(\bm{y_{i}})).
\end{equation}
Hence, on one hand, the Neural Network is trained using as input the design matrix $\bm{Y}$ with rows equal to the vectors of covariates $\bm{y}_i$ of all patients $i=1,...,n$, and the posterior probabilities $\{\hat{u}_{i}=\mathbb{E}[z_{i}|\bm{x}_{i},t_{i},\delta_{i}]\}_{i=1}^{n}$ as the response variable vector. A single hidden layer feedforward Neural Network is defined in the following way
\begin{equation*}
    \theta(\bm{y}) :=\sigma\left(\sum_{k=1}^{\text{\#hidden neurons}}c_{k}\sigma(\bm{a}_{k}^{'}\bm{y}) + c_{0}\right),
\end{equation*}
where $c_{k} \in \mathbb{R}$ and $\bm{a}_{k} \in \mathbb{R}^{q+1}$ $\forall k$ are unknown weights, and $\sigma$ is the logistic activation function. The weights of the Neural Network are determined maximizing (\ref{eq:negcrossentr}) with the quasi-Newton method \textit{BFGS} \cite{flet87}, which also provides an estimate of the Hessian matrix evaluated at the estimated parameters. The GAM, on the other hand, is a flexible extension of Generalized Linear Models (GLM) that represents the relationship between a response variable and the covariates as a sum of smooth functions. These smooth effects are commonly estimated using smoothing splines by optimizing equation (\ref{eq:negcrossentr}) with a penalty on each spline term to control excessive complexity. Estimation is typically carried out using the local scoring algorithm, which iteratively fits weighted additive models via backfitting \cite{hastie1986}.

Lastly, $\bm{\gamma}$ and $\bm{\xi}$ are updated maximizing
\begin{equation*}
\label{optim:gammaxi}
    \ell_{c}(\bm{\gamma},\bm{\xi} \, ; \, \bm{X},\bm{t},\bm{\delta}, \bm{u})\bigg|_{\theta(\cdot)=\hat{\theta}(\cdot),\; \bm{u}=\bm{\hat{u}}}
\end{equation*}
using \textit{Nelder-Mead} simplex numerical optimization method \cite{nelder1965}. Thus, our proposed EM algorithm starts from an initial guess $\bm{u}^{(0)},\theta^{(0)},\bm{\gamma}^{(0)},\bm{\xi}^{(0)}$ and at iteration k+1 computes $\bm{u}^{(k+1)},\theta^{(k+1)},\bm{\gamma}^{(k+1)},\bm{\xi}^{(k+1)}$ using the parameters values computed during the previous iteration, i.e. $\bm{u}^{(k)},\theta^{(k)},\bm{\gamma}^{(k)},\bm{\xi}^{(k)}$, through the just described procedure. The algorithm stops when the difference of the log-likelihood (\ref{log-likelihood}) between two consecutive iterations is below a given threshold $\epsilon$ fixed beforehand. 

\section{Simulation Study}
\label{sec:sim}
We conducted a simulation study \cite{morris2019} to compare the consistency and efficiency of our proposed model with those of the standard Weibull mixture cure model \cite{lambert2007} in estimating the fraction of cured individuals across scenarios with different numbers of covariates and different functional specifications of the cure fraction.

The data were simulated from the mixture cure model in Equation (\ref{eq:overallNS}). In particular, NS of fatal patients was simulated from a Weibull distribution to achieve short survival for the uncured individuals, while the ES $S_0$ was simulated from a Weibull distribution such that the cured group had a long life expectancy. We considered four sample sizes ($n = 100, 500, 1{,}000, 10{,}000$), a follow-up period of $\tau = 30$ years, and five variables: \textit{age} (proxy for age at diagnosis), \textit{period} (proxy for period of diagnosis), \textit{sex}, \textit{socioeconomic} (proxy for socioeconomic status), and  \textit{logtms} (proxy for the natural logarithm of the tumor size). \textit{age} was simulated from a Poisson distribution with mean 45, shifted by 20; for the variable \textit{period}, $\frac{n}{3}$ observations were simulated from a uniform distribution on the interval $[0, \frac{\tau}{2}]$, and the remaining observations from the interval $[\frac{\tau}{2}, \tau]$; \textit{sex} was simulated from a Bernoulli distribution with parameter 0.48. The variable \textit{socioeconomic} was simulated from a multinomial distribution with probability vector $(0.3, 0.6, 0.1)$ (low, medium, high), which was then converted into three dummy variables, \textit{socioeconomic\_low}, \textit{socioeconomic\_med}, and \textit{socioeconomic\_high}. Finally, \textit{logtms} was computed as the natural logarithm of a value simulated from a chi-squared distribution with two degrees of freedom. We considered three different specification of the true cure probability, each of which generated a different simulation scenario: 
\begin{itemize}
    \item[\textbf{1.}] Linear and non-linear effects and interactions of three variables: 
        \begin{equation} 
            \begin{split} 
                \theta_{0}(\bm{x})= & \; \text{logistic}[0.5-2.5 \cdot age + 0.1 \cdot sex - 2.8 \cdot (age)^2 + \\ 
                & + 0.8 (period)^2 + 1.2 \cdot (|period| \cdot age)].
            \end{split}
        \end{equation}
    \item[\textbf{2.}] Linear and non-linear effects and interactions of seven variables:
    \begin{equation}
            \begin{split}
                \theta_{0}(\bm{x})= & \; \text{logistic}[0.5-2.5 \cdot age + 0.1 \cdot sex - 2.8 \cdot (age)^2 + \\
                & + 0.8 (period)^2 + 1.2 \cdot (|period| \cdot age)) \\
                & - 0.1 \cdot socioeconomic\_low + 0.01 \cdot socioeconomic\_mid \\
                & + 0.1 \cdot socioeconomic\_high - 0.1 \cdot (logtms \cdot age)].
            \end{split}
        \end{equation}
    \item[\textbf{3.}] Linear effects of seven variables:
    \begin{equation}
            \begin{split}
                \theta_{0}(\bm{x})= & \; \text{logistic}[0.5-2.5 \cdot age + 0.1 \cdot sex + 0.02 \cdot period + \\ 
                & - 0.1 \cdot socioeconomic\_low + 0.01 \cdot socioeconomic\_mid + \\
                & + 0.1 \cdot socioeconomic\_high - 0.1 \cdot logtms].
            \end{split}
        \end{equation}
\end{itemize}
For each sample size, we simulated $250$ datasets and chose the cure fraction $\theta_0$ as the estimand of interest. 

We compared three methods derived from the mixture cure model in (\ref{eq:overallNS}), which differ in how the non-parametric component has been specified: a GAM with smooth interactions and a logit link, a Neural Network with a logistic activation function, and a GLM with a logit link (proposed by Lambert et al (2007) \cite{lambert2007}). We implemented the simulation in R \cite{r}, and further details and code can be found at the following link: \hyperlink{https://github.com/fabradimitra/Flexible-in-RS-cure-model-simulation}{\textit{https://github.com/fabradimitra/Flexible-in-RS-cure-model-simulation}}. 

The estimation of the non-parametric component for the GAM model was implemented using the \textit{gam} function of the package \textit{mgcv} \cite{wood2011}. As briefly stated in the previous section, this function fits a GAM by penalized likelihood maximization, in which the model negative log-likelihood is modified by the addition of a penalty for each smooth function, penalizing “wiggly”  behaviors. The trade-off between roughness and wiggliness is controlled by a smoothing parameter which is estimated using Generalized Cross-Validation (GCV). For Scenario 1, we considered a main effect for \textit{sex}, a smooth function of \textit{age} and \textit{period}, and a smooth tensor product interaction between \textit{age} and \textit{period}. For Scenarios 2 and 3, we considered the effects of Scenario 1 and added main effects for \textit{socioeconomic\_low}, \textit{socioeconomic\_mid}, and \textit{socioeconomic\_high}, as well as smooth tensor product interactions between \textit{age} and \textit{logtms}, and between \textit{period} and \textit{logtms}. The estimation of the non-parametric component for the Neural Network was instead implemented with the package \textit{nnet} \cite{nnet}. Since we wanted to explore to which extent increasing the width of the network enhances the accuracy of cure probability estimation, we set the layer size to $4, 8,$ and $16$ neurons ($\text{Nnet}_{4}$, $\text{Nnet}_{8}$, $\text{Nnet}_{16}$ models respectively) and considered all available variables in the specific scenario (three variables for Scenario 1 and seven variables for Scenarios 2 and 3). Last but not least, the GLM was implemented using the function \textit{glm} of the \textit{stats} package. For Scenario 1, we considered main effects of \textit{sex}, \textit{age} and \textit{period}. For Scenarios 2 and 3, we considered main effects of \textit{sex}, \textit{age}, \textit{period}, \textit{socioeconomic\_low}, \textit{socioeconomic\_mid}, \textit{socioeconomic\_high}, and \textit{logtms}. 

\begin{figure}[!ht]
  \centering
  \includegraphics[width=\textwidth]{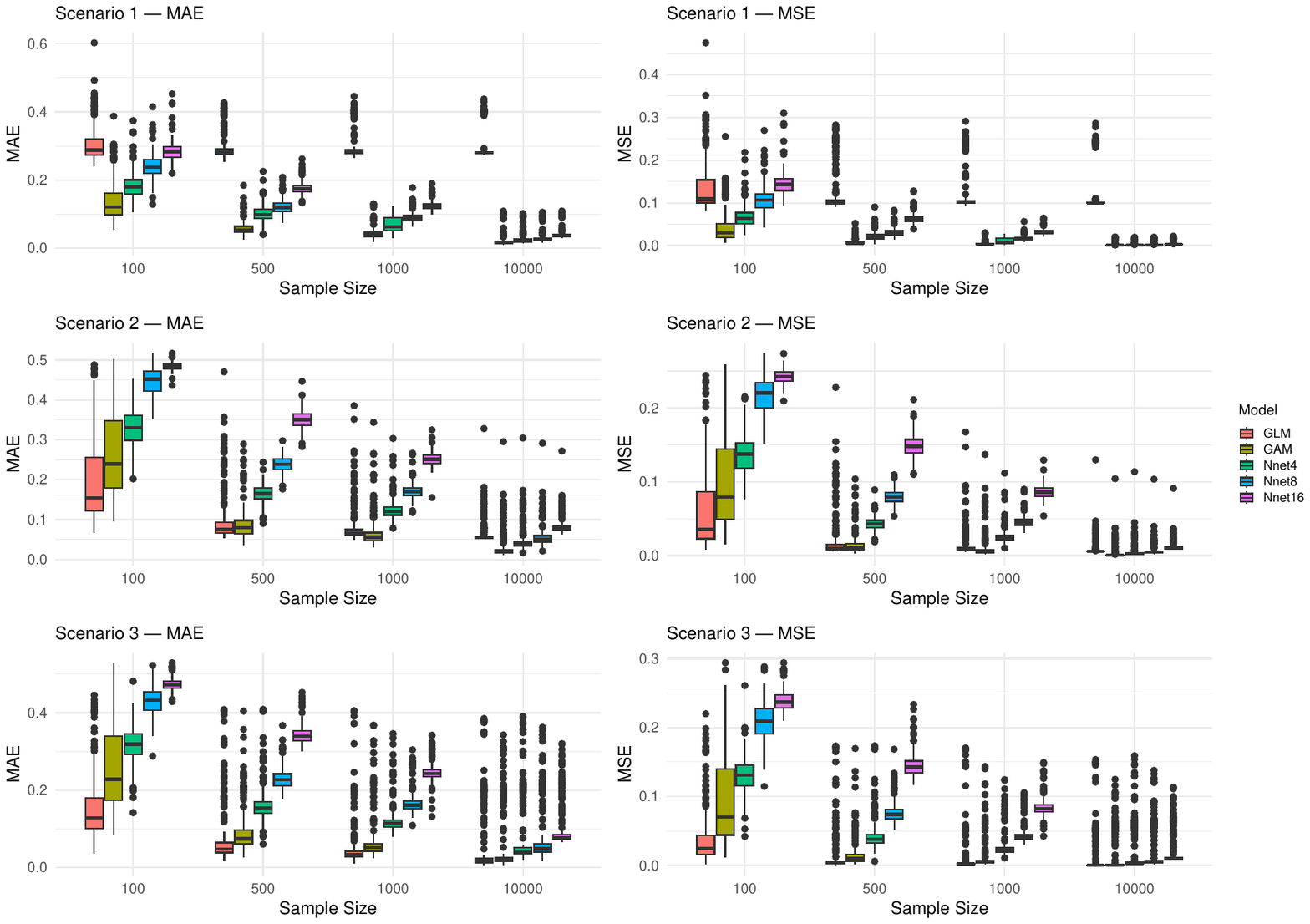} 
  \caption{Boxplots of the distributions of the MAE and MSE estimates of the non-parametric component for the GAM model, the GLM model, and the Neural Network models with $4, 8$, and $16$ neurons for each simulated Scenario. Results from a simulation study.}
  \label{fig:boxplots}
\end{figure}

We standardized all covariates except the dichotomous covariates prior to simulating the datasets and running the estimation algorithms, parallelizing the computations across 16 workers. While the GLM model and the Nnet models completed their computations in less than an hour, depending on the scenario, the GAM model was the most time-consuming, completing its computations in between 10 and 20 hours. In Figure \ref{fig:boxplots} we reported the distribution of Mean Absolute Error (MAE) (left column) and Mean Squared Error (MSE) (right column) for all considered models.

As sample size increases, the proximity of these distributions toward zero provides evidence of consistency, while their relative magnitudes across models quantify efficiency, with smaller MAE or MSE indicating more accurate and stable estimates in finite samples. When we considered three covariates (Scenario 1), already with a small sample size of $100$, our proposed models had lower MAE and MSE than the GLM model, with the GAM model performing consistently better than all other models for all sample sizes. When we increased the number of covariates, we observed that more observations were needed to obtain comparable results with the standard model. In particular, when the true cure fraction contained many interactions (Scenario 2), the GAM model started performing better than the standard GLM model from $1{,}000$ observations, followed by the other proposed models at $10{,}000$ observations. For this latter sample size, all our proposed models except for $\text{Nnet}_{16}$ yielded better results than the GLM model, particularly with respect to MAE, which is less sensible to outliers than MSE. When there were no interactions in the true cure fraction and only linear effects (Scenario 3), the GLM model had both lower MAE and MSE for all sample sizes except for $10{,}000$ observations, for which the GAM model showed comparable results.
\section{Assessing survival trajectories of colon cancer patients}
\label{sec:appl}
The data considered in the application were the individual colon cancer data collected by the population-based cancer registry of Varese in Italy. The cohort consisted of all 8,647 patients residing in the municipality of Varese who were diagnosed with colon cancer from 1993 to 2013. The follow-up time stopped the $31^{\text{st}}$ December 2013. According to the \textit{World Health Organization} colon cancer is one of the most frequent cancer in the population. Moreover, the cure plateau is visible in a reasonable follow-up period, generally around $10$ years \cite{lambert2007,lambert2009}, although extending the follow-up time could further validate the cure assumption. The covariates under consideration were the \textit{age at diagnosis} and the \textit{period of diagnosis}, as these variables stand out as the ones mostly affecting the outcome of colon cancer survival, while \textit{sex} seems to be an independent prognostic factor for this type of cancer \cite{cheung2013,samawi2018}. 

Building on the results of the simulation study in the previous section, in this application we compared three models: the GLM model, the GAM model and the Neural Network model with $4$ hidden neurons. We implemented the models with the same tools as we did in the simulation study, while the ES was computed with the function \textit{survexp} of the R-package \textit{survival} \cite{survivalpackage} provided with the population life tables. The predictor in the GAM is the same as the one considered for the Scenario 1 of the simulation study in Section \ref{sec:sim}, but without the term involving the \textit{sex} covariate. 

To facilitate the reporting of results, we have subdivided age into five classes: 0-44, 45-54, 55-64, 65-74, 75-99; and \textit{period} into four periods: 1993-1997, 1998-2002, 2003-2007, and 2008-2013. 

\begin{table}[!ht]
\centering
\caption{Summary of age at diagnosis and follow-up events by age group and period group, with median (first quartile - fourth quartile) age and n/N (\%) of events.}
\label{2tab:dataset_analysis}
\small % Slightly smaller font to fit the width comfortably
\begin{tabular}{lcccc}
%\toprule
 & \multicolumn{4}{c}{\textbf{Period group}} \\
\cmidrule(lr){2-5}
 & \textbf{[1993, 1998)} & \textbf{[1998, 2003)} & \textbf{[2003, 2008)} & \textbf{[2008, 2013]} \\
\midrule
\multicolumn{5}{l}{\textbf{Age group: [0, 55) (N = 826)}} \\
\quad Age & 49 (44--52) & 50 (45--52) & 50 (45--52) & 50 (45--52) \\
\quad Events & 94 / 185 (51\%) & 90 / 213 (42\%) & 81 / 222 (36\%) & 57 / 206 (28\%) \\
\midrule
\multicolumn{5}{l}{\textbf{Age group: [55, 65) (N = 1700)}} \\
\quad Age & 60 (57--63) & 60 (58--63) & 60.5 (58--63) & 60 (58--63) \\
\quad Events & 220 / 377 (58\%) & 201 / 411 (49\%) & 193 / 518 (37\%) & 92 / 394 (23\%) \\
\midrule
\multicolumn{5}{l}{\textbf{Age group: [65, 75) (N = 1000)}} \\ % Note: Corrected typo in image which said [65,65)
\quad Age & 70 (67--72) & 70 (67--72) & 70 (67--72) & 70 (67--72) \\
\quad Events & 439 / 570 (77\%) & 438 / 660 (66\%) & 402 / 811 (50\%) & 278 / 753 (37\%) \\
\midrule
\multicolumn{5}{l}{\textbf{Age group: [75, 99] (N = 3394)}} \\
\quad Age & 81 (78--85) & 81 (77--85) & 81 (78--84) & 81 (78--85) \\
\quad Events & 611 / 632 (97\%) & 671 / 758 (89\%) & 676 / 884 (76\%) & 667 / 1,053 (63\%) \\
%\bottomrule
\end{tabular}
\end{table}

In Table \ref{2tab:dataset_analysis}, we reported summary statistics for the discretized variables \textit{age at diagnosis} and \textit{period of diagnosis}. The table consists of four stacked sub-tables, each referring to an age class identified by a label that also reports the number of observations belonging to that age group. The columns of each sub-table correspond to period classes; thus, each cell of Table \ref{2tab:dataset_analysis} reports information for the corresponding age-period combination. In particular, the row \textit{Age} reports the median age at diagnosis, with the first and third quartiles shown in brackets, while the row \textit{Events} reports the number of observed events over the total number of observations, with the corresponding percentage in parentheses. For each age class, we observe a fairly homogeneous distribution of observations across periods, along with a decrease in the number of events in more recent periods. This decrease is most likely due to administrative censoring, as individuals in later periods are observed for shorter follow-up times. The most numerous age class is the oldest, as expected given that cancer primarily affects older individuals. This age class is also the one at highest risk of death, as indicated by the higher percentages of observed events compared with the other age classes. The first and third quartiles show that, in the youngest age class, 25\% of individuals were younger than $45$ years in each age-period combination, and analogously, in the oldest age class, 25\% of individuals were diagnosed at ages above $85$ years.

\begin{figure}[!ht]
  \centering
  \includegraphics[width=\textwidth]{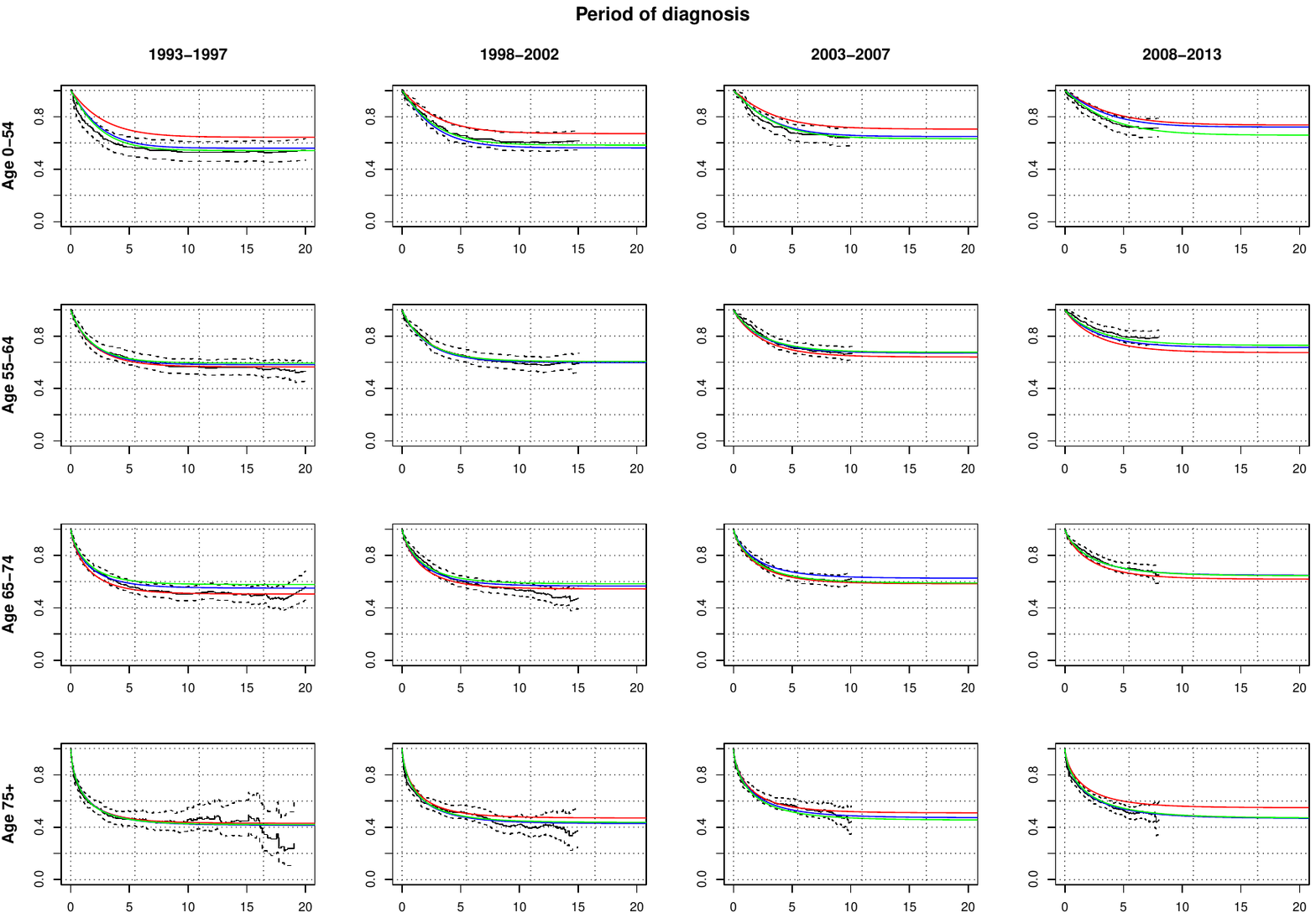} 
  \caption{net survival (NS) of colon cancer patients estimated by the Ederer II life-time method \cite{ederer1961} (black line) with 95\% confidence interval, by the GLM model (red line), by the GAM model (blue line), and by the Neural Network  model with 4 hidden neurons (green line). Y-axis represents the NS, while the x-axis represents the follow-up time expressed in years.}
  \label{GLM_vs_GAM_vs_NN}
\end{figure}

Figure \ref{GLM_vs_GAM_vs_NN} shows the NS functions estimated for different age-period combinations. The black line represents the Ederer II non-parametric estimator \cite{ederer1961}, computed through the R package \textit{relsurv} \cite{relsurv}, with associated confidence intervals (dashed black lines). The red line is the estimated GLM model, while the blue line is the GAM model. Finally, the green line is the Neural Network model with 4 hidden neurons. The colored curves were obtained by estimating the survival function for all individuals belonging to each age-period class, and then averaging the results in each time-point. Although the non-parametric estimates should not be regarded as the true values of NS, we expect the estimated cure fraction to be as close as possible to the point at which these estimates appear to reach an asymptote. Overlaying the model-based estimates on the non-parametric estimates has already been suggested elsewhere \cite{poharperme2016}. The figure shows stronger agreement between the non-parametric estimates and the more flexible models. In contrast, the GLM model often fails to fall within the confidence intervals of the non-parametric estimates.

\begin{figure}[!ht]
  \centering
  \includegraphics[width=\textwidth]{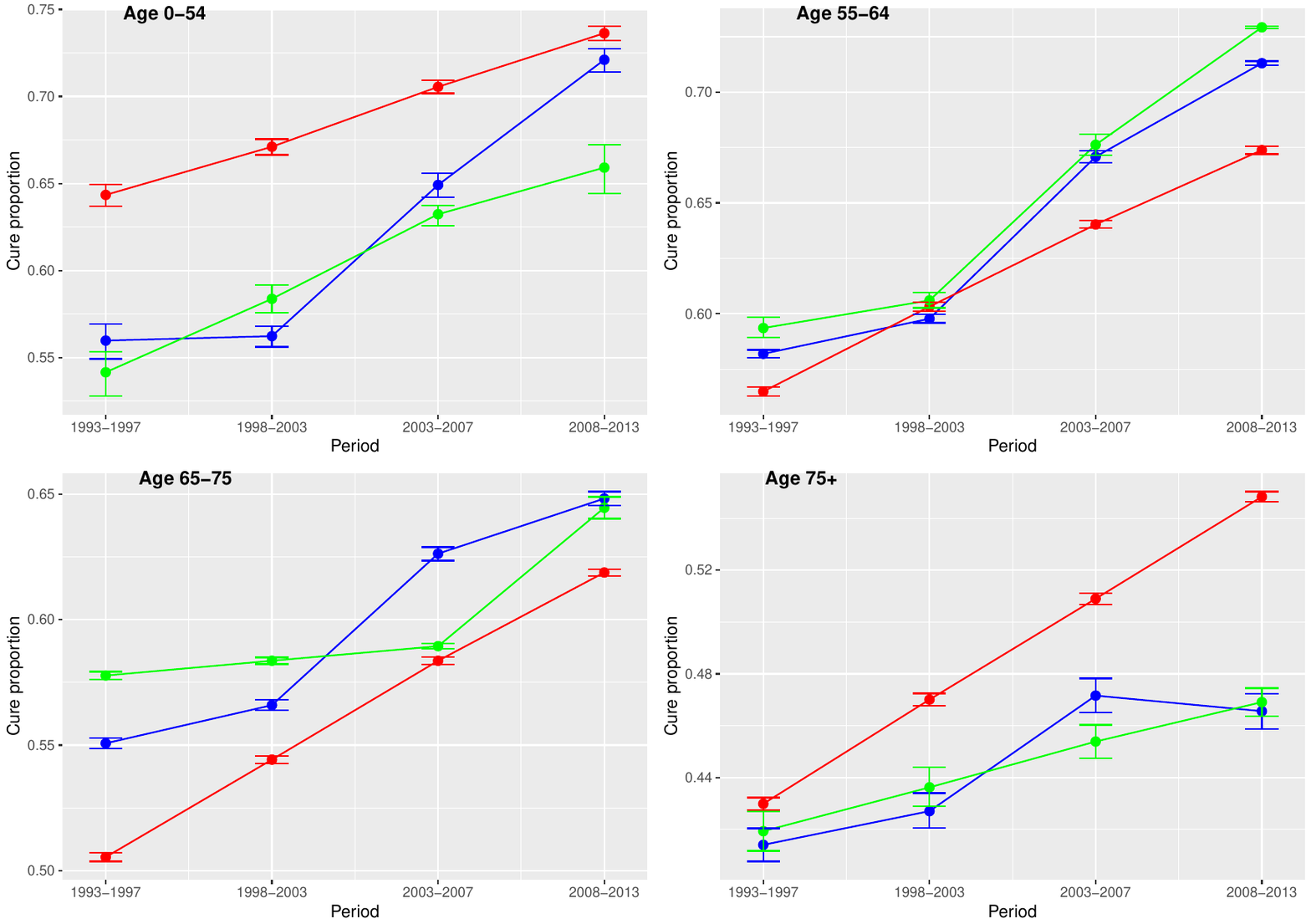} 
  \caption{Estimated cure proportion by period of diagnosis for each age class with 95\% confidence intervals: the GLM mixture model (red line), the GAM mixture cure model (blue line), and the Neural Network mixture cure model (green line).}
  \label{cure_proportion_comparison}
\end{figure}

Figure \ref{cure_proportion_comparison} shows the trends of the estimated cure proportions by period of diagnosis for each age class, with 95\% confidence intervals computed using bootstrap with $5000$ replicates \cite{efron1987} through the R package \textit{boot} \cite{boot}. In this figure we notice the different trends between the flexible models and the GLM model. The latter model can only fit a linear trend for the cure proportion across periods, while the former models, which consider nonlinear effects and interactions between the two covariates, allow for a more flexible growth or decline in the predicted cure proportions.

\begin{table}[!ht]
\caption{Parameters' estimates, associated standard errors and associated p-values of the Wald test for the three considered models: the GLM model, the model GAM, and the Neural Network model with 4 hidden neurons.}
\label{param_est}
\begin{adjustbox}{max width=\textwidth}
\begin{tabular}{lcccccc}
%\hline
Parameters                &                      &                      &                      &                      &                      &                      \\
                          & GLM                  & Wald test            & GAM                  & Wald test            & $\text{Nnet}_{4}$    & Wald test            \\ \hline
\textit{Cure fraction}             &                      &                      &                      &                      &                      &                      \\
Intercept                 & 0.262 (0.039)        & p$<$0.05               & \textbf{-}                    & \textbf{-}                    & \textbf{-}                    & \textbf{-}                    \\
Age                       & -0.292 (0.037)       & p$<$0.05               & \textbf{-}                    & \textbf{-}                    & \textbf{-}                    & \textbf{-}                    \\
Period of diagnosis       & 0.179 (0.038)        & p$<$0.05               & \textbf{-}                    & \textbf{-}                    & \textbf{-}                    & \textbf{-}                    \\
                          &                      &                      &                      &                      &                      &                      \\ \hline
\textit{Scale parameter} $\lambda$ &                      &                      &                      &                      &                      &                      \\
Intercept                 & -6.533 (0.051)       & p$<$0.05               & -6.569 (0.035)       & p$<$0.05               & -6.576 (0.036)       & p$<$0.05               \\
Age                       & 0.287 (0.048)        & p$<$0.05               & 0.223 (0.033)        & p$<$0.05               & 0.233 (0.033)        & p$<$0.05               \\
Period of diagnosis       & -0.131 (0.048)       & p$<$0.05               & -0.144 (0.035)       & p$<$0.05               & -0.171 (0.035)       & p$<$0.05               \\
                          &                      &                      &                      &                      &                      &                      \\ \hline
\textit{Shape parameter} $\beta$   &                      &                      &                      &                      &                      &                      \\
Intercept                 & -0.205 (0.020)       & p$<$0.05               & -0.207 (0.018)       & p$<$0.05               & -0.211 (0.018)       & p$<$0.05               \\
Age                       & -0.147 (0.021)       & p$<$0.05               & -0.162 (0.017)       & p$<$0.05               & -0.158 (0.017)       & p$<$0.05               \\
Period of diagnosis       & 0.012 (0.020)       & p=0.541              & 0.011 (0.018)        & p=0.543              & 0.003 (0.018)        & p=0.878              \\
                          & \multicolumn{1}{l}{} & \multicolumn{1}{l}{} & \multicolumn{1}{l}{} & \multicolumn{1}{l}{} & \multicolumn{1}{l}{} & \multicolumn{1}{l}{} \\ 
                          %\hline
\end{tabular}
\end{adjustbox}
\end{table}

Table \ref{param_est} shows the parameter estimates of the three models considered, along with the estimated standard errors and the p-values of the Wald test for the significance of the estimated parameters. The models all agree on the non-significance of the regression parameter associated with the covariate \textit{period of diagnosis} in the shape parameter $\beta$. Similarly, they all confirm the significance of all other regression parameters in both the scale $\lambda$ and shape $\beta$ parameters. The models do not show significant differences in the estimated values of the parameters for the NS of fatal cases. Instead, their primary distinction lies in the estimated asymptote of the NS function, i.e., the cure fraction.

We then selected the model that best describe new data by evaluating its out-of-sample log-likelihood. For this purpose, we conducted 5-fold cross-validation, wherein we fit the models to four subsets of the dataset each time and then measure the log-likelihood in the holdout subset. By averaging the results in the holdout subsets, we obtained a measure of the generalization performance of our models. Even though computationally expensive, this procedure is known to give robust estimate of the generalization performance of the models at hand. We also considered the \textit{Akaike Information Criterion} (AIC) \cite{akaike1973} and the \textit{Bayesian Information Criterion} (BIC) \cite{schwarz1978} to assess the trade-off between goodness of fit and model complexity of the considered models. The number of degrees of freedom in the GLM is given by the number of covariates, while the GAM degrees of freedom are calculated by adding up the degrees of freedom used by the parametric and non-parametric (smooth) terms in the model. Things get a bit more complicated for the Neural Network. While one may think that the generalization performance of Neural Networks depend on the total number of network parameters, Bartlett (1998) \cite{bartlett1998} showed that it actually depends on the $L_{1}$ norm between the weights connecting the hidden layer to the output layer rather than the total number of weights in the model. Using projection theory of linear models, Ingrassia and Morlini (2005) \cite{ingrassia2005}, derived an equivalent number of degrees of freedom in a one-hidden-layer network to be used in model selection criteria: the number of hidden-to-output weights in the network. They also linked this number to the $L_{1}$ norm of the input-to-output weights, showing agreement with the theorem proved by Bartlett (1998) \cite{bartlett1998}. In our case, the number of hidden units was set to 4; therefore, considering also the bias parameter, the equivalent number of model degrees of freedom was taken as 5. As a result, the overall degrees of freedom for the three models were 6, 16.8 and 9 for the GLM model, the GAM model, and the Neural Network model, respectively.

\begin{table}[!ht]
\centering
\caption{AIC, BIC, and average 5-fold cross-validation log-likelihood for the three considered models.
}
\begin{tabular}{lccc}
%\hline
Model selection          & \multicolumn{1}{l}{} & \multicolumn{1}{l}{} & \multicolumn{1}{l}{} \\
                         & GLM                  & GAM                  & $\text{Nnet}_{4}$    \\ \hline
Model selection scores &                      &                      &                      \\
AIC                      & $90,700.34$          & $90,620.64$          & $\bm{90,607.83}$     \\
BIC                      & $90,742.73$     & $90,737.70$          & $\bm{90,671.41}$          \\
                         &                      &                      &                      \\ \hline
5-fold cross-validation  &                      &                      &                      \\
Average log-likelihood   & $-9,071.2$             & $-\bm{9,063.6}$            & $-9,066.9$        \\
                         & \multicolumn{1}{l}{} & \multicolumn{1}{l}{} & \multicolumn{1}{l}{} \\ 
%hline                         
\end{tabular}
\label{model_selection}
\end{table}

Table \ref{model_selection} shows the AIC and BIC scores, as well as the average log-likelihood from the 5-fold cross-validation for the three models considered. All the metrics agree that the more flexible models perform better than the GLM model. Specifically, the 5-fold cross-validation highlights the superior generalization performance of the GAM model compared to the others, while the goodness-of-fit measures emphasize the better adaptation of the Neural Network. However, since the criteria for counting the degrees of freedom differ among the three models, we will rely more on the results of the generalization measure for model selection. Therefore, in the subsequent analysis, we will focus exclusively on the GAM model.

\begin{table}
\setlength{\tabcolsep}{8pt}
\centering
\caption{Estimates of the proportion of cured cases and median survival time of fatal cases with corresponding 95\% confidence intervals, obtained from the model using a GAM to estimate the non-parametric component.}
\resizebox{\textwidth}{!}{\begin{tabular}{l l l l l} % S column type from siunitx
%\toprule
%\midrule
\multicolumn{1}{l}{Age class} & \multicolumn{4}{c}{Period of diagnosis} \\
\multicolumn{1}{l}{}  & {1993-1997} & {1998-2002} & {2003-2007} & {2008-2013}\\
\midrule
\multicolumn{5}{l}{\textit{Proportion of cured}}\\
0-54 & 0.560 (0.549-0.569) & 0.562 (0.556-0.568) & 0.649 (0.642-0.656) & 0.721 (0.714-0.727)\\
55-64 & 0.582 (0.580-0.584) & 0.598 (0.596-0.600) & 0.671 (0.668-0.674) & 0.713 (0.712-0.714)\\
65-74 & 0.551 (0.549-0.553) & 0.566 (0.564-0.568) & 0.626 (0.624-0.629) & 0.648 (0.645-0.651)\\
75+ & 0.414 (0.408-0.420) & 0.427 (0.420-0.434) & 0.472 (0.465-0.478) & 0.466 (0.459-0.472)\\
\midrule
\multicolumn{5}{l}{\textit{\textit{Median survival time for fatal cases (years)}}} \\
0-54 & 1.825 (1.765-1.878) & 1.994 (1.950-2.035) & 2.271 (2.228-2.310) & 2.591 (2.527-2.649)\\
55-64 & 1.302 (1.291-1.312) & 1.485 (1.472-1.497) & 1.691 (1.680-1.703) & 1.915 (1.900-1.930)\\
65-74 & 1.024 (1.017-1.031) & 1.166 (1.159-1.173) & 1.333 (1.325-1.341) & 1.506 (1.497-1.514)\\
75+ & 0.750 (0.742-0.757) &  0.861 (0.853-0.870) & 0.981 (0.973-0.989) & 1.120 (1.111-1.129)\\
%\toprule
\end{tabular}}
\label{model_results}
\end{table}

Table \ref{model_results} shows the estimated proportion of cured cases and the estimated median survival time for the uncured group, along with 95\% confidence intervals calculated using bootstrap, for each combination of \textit{age} and \textit{period} of diagnosis. The median survival time $t_{0.5}$ of a Weibull random variable can be easily calculated as $t_{0.5}=\frac{1}{\lambda} \cdot (\text{log}2)^{\frac{1}{\beta}}$, where $\lambda$ and $\beta$ are the scale and the shape parameters of the Weibull distribution, respectively. There is a considerable improvement in survival across the investigated time periods which is consistent with the diagnostic and therapeutic improvement over the years. As expected, younger individuals have higher odds of being cured than the elderly. 

A related quantity to the median survival time is the time to cure $\tilde{t}_{\alpha}$, i.e. the time after which only a negligible proportion $\alpha$ of fatal cases still survives. If $\alpha$ is sufficiently small (say 0.05 or 0.01), survivors at $\tilde{t}_{\alpha}$ could be practically considered as cured. When we are dealing with a Weibull distribution, this quantity can be calculated as $\tilde{t}_{\alpha}=\frac{1}{\lambda} \cdot \text{log}(\alpha^{\frac{1}{\beta}}).$

\begin{table}[!ht]
\setlength{\tabcolsep}{8pt}
\centering
\caption{Estimates of time to cure (in years), i.e. time at which only 5\% and 1\% of fatal individuals survive ($\tilde{t}_{0.05}$ and $\tilde{t}_{0.01}$ respectively), obtained from the model using a GAM to estimate the non-parametric component.}
\resizebox{\textwidth}{!}{\begin{tabular}{l c c c c} % S column type from siunitx
%\midrule
\multicolumn{1}{l}{Age class} & \multicolumn{4}{c}{Period of diagnosis} \\
\multicolumn{1}{l}{}  & {1993-1997} & {1998-2002} & {2003-2007} & {2008-2013}\\
\midrule
\multicolumn{5}{l}{$\tilde{t}_{0.05}$}\\
0-54 &  6.71 (6.66-6.76) &  7.49 (7.44-7.54) & 8.46 (8.41-8.51) & 9.49 (9.43-9.55)\\
55-64 &  6.26 (6.24-6.28) &  7.05 (7.03- 7.08) &  7.96 (7.94-7.99) &  8.88 (8.85-8.92)\\
65-74 &  5.97 (5.95-5.99) &  6.71 (6.69-6.73) &  7.56 (7.54-7.58) &  8.46 (8.43-8.48)\\
75+ &  5.58 (5.56-5.60) &  6.30 (6.28-6.32) &  7.08 (7.06-7.10) &  7.98 (7.96-8.00)\\
\midrule
\multicolumn{5}{l}{\textit{$\tilde{t}_{0.01}$}} \\
0-54 & 10.32 (10.24-10.40) & 11.52 (11.44-11.60) & 13.01 (12.93-13.08) & 14.59 (14.49-14.68)\\
55-64 & 9.62 (9.59-9.65) & 10.85 (10.81-10.88) & 12.24 (12.20-12.28) & 13.66 (13.60-13.71)\\
65-74 &  9.17 (9.15-9.20) &  10.31 (10.29-10.34) & 11.63 (11.60-11.66) & 13.00 (12.97-12.03)\\
75+ &  8.58 (8.55-8.61) &  9.68 (9.65-9.71) &  10.89 (10.86-10.92) &  12.27 (12.24-12.30)\\
\end{tabular}}
\label{time_to_cure}
\end{table}

Table \ref{time_to_cure} shows the estimated time at which only 1\% and 5\% of fatal cases survive. We observe, on one hand, a decreasing trend across age classes, which is not surprising since older individuals with fatal conditions tend to die early in the follow-up period. On the other hand, we observe a positive trend across periods, indicating improvements in medical treatment. An individual diagnosed in later periods must wait longer to be considered cured due to the overall improvement in survival outcomes for colon cancer. Insurance companies can calculate the risk level for different $\alpha$ values and then choose the time to cure, $\tilde{t}_{\alpha}$, at which to grant insurance policies to cured individuals.
\section{Discussion}
\label{sec:conc}
More than half of cancer patients in Europe survive for 5 years or more after diagnosis, and current estimates indicate that there are over 23.7 million cancer survivors in Europe today \cite{deangelis2024}. Hence, partitioning the patients population into two distinct groups, those having the same risk as the general population and those facing an excess risk due to the disease, is a reasonable assumption from a biological perspective. Mixture models allow simultaneous estimation of the probability of cure and the survival outcomes of fatal individuals. However, standard RS Weibull cure models \cite{deangelis1999,lambert2007} assume that the relationship between the cure probability and the covariates is linear on the logit scale. 

In this paper, we proposed to use of more flexible functions to link covariates to the cure probability in order to protect against model misspecification. In particular, we considered GAMs with smooth interactions and one-hidden layer feedforward Neural Networks. The epidemiological plausibility of these approaches stems from medical advancements, which do not always induce linear effects of the period of diagnosis, age at diagnosis, and other factors on the cure fraction. There may be sudden shifts in cure fractions from one period to another due to the introduction of new treatments, and the same treatment may be effective for younger age groups but not for older ones. 

After describing the methodology, we derived an EM algorithm for this class of models to obtain maximum likelihood estimates of the model parameters.

We conducted a simulation study with three different data-generating scenarios, differentiated by how the cure fraction was specified as a function of the covariates, and we measured the performance of each considered model by means of the MAE and MSE between the estimated and true cure proportions (cf. Figure \ref{fig:boxplots}). Specifying the cure fraction as a function of linear and non-linear effects, as well as interactions, among three covariates, yielded better results for our proposed Neural Network models (regardless of the size of the hidden layer) and for the GAM model compared to the standard GLM model. When we increased the number of covariates while still including non-linear effects and interactions, we observed that the proposed models required more observations, but eventually caught up with and surpassed the standard GLM model when the sample size reached $10{,}000$ observations. When our proposed models were misspecified and the GLM model was well specified, we observed the GLM model yielding the best results across all sample sizes. However, for a sample size of $10{,}000$, we observed a greater relative difference in the distributions between the GLM model and the GAM model when the GLM model was misspecified than when the GAM model was misspecified. This suggests that, when the sample size is sufficiently large, it is still worthwhile to use the GAM model even if a linear predictor is believed to be sufficient and the number of covariates are more than the ones usually considered in population-based studies, namely \textit{age at diangosis}, \textit{period of diagnosis}, and \textit{sex}. The same cannot be said for the Neural Network models, for which the relative differences in the distributions with respect to the GLM model were similar when either model was misspecified.

We applied our proposed methodology to a cohort of colon cancer patients diagnosed in the municipality of Varese between 1993 and 2013. The more flexible models showed better agreement with the Ederer II non-parametric estimates than the standard GLM model, even for the oldest age group, which is always the most challenging to fit. Notably, these models not only performed better for the youngest age group, which is the least numerous, but also adapted more effectively to classes with shorter follow-up periods. In contrast, the GLM model generally showed the opposite pattern. 

We found that all the estimated models differed only in the nonparametric component, while yielding almost identical results for the parameters describing the survival time of fatal cases (cf. Table \ref{param_est}). This is unsurprising, as the NS curves within each age-period category showed similar shapes but differed in their vertical positioning (cf. Figure \ref{GLM_vs_GAM_vs_NN}). Furthermore, the flexible models showed greater concordance in their cure fraction trends, whereas the standard model was constrained to follow a specific trajectory over diagnosis period (cf. Figure \ref{cure_proportion_comparison}). Overall, with relatively few choices required, the proposed approaches captured nonlinear trends in the cure proportion across different time periods and age classes by accounting for nonlinear main effects and interactions in the specification of the cure fraction.

Our proposed models performed better than the GLM model according to AIC, BIC, and the out-of-sample log-likelihood computed using 5-fold cross-validation. Among the models considered, the GAM demonstrated the best out-of-sample performance and was therefore selected as the best candidate model. We did not rely too much on information criteria because the degrees of freedom were calculated differently across the models, and in cure models the primary interest lies in the tail of the distribution. AIC and BIC tend to favor models that fit the early follow-up period well, as most events occur early, but may perform poorly in identifying the cure model that best estimates the cure proportion. By contrast, the out-of-sample log-likelihood is better suited for comparing cure models, as it directly reflects performance on unobserved data and thus the ability to generalize, including beyond the observed period. Nevertheless, since the experimental study in Di Mari et al. (2025) \cite{DiMari2025} was conducted on the same cohort of patients considered in this work and relied on AIC and BIC to compare the proposed approach with the cure model of De Angelis et al. (1999) \cite{deangelis1999}, a retrospective comparison of goodness of fit between the model proposed in Di Mari et al. (2025) \cite{DiMari2025} and the models proposed in this paper is still possible. This comparison can be done by examining the results reported in Table 1 of Di Mari et al. (2025) \cite{DiMari2025} and Table \ref{model_selection} in Section \ref{sec:appl} of this work. Based on these results, increasing the number of Weibull mixture components in the model proposed by De Angelis et al. (1999) \cite{deangelis1999} (as proposed in Di Mari et al. (2025) \cite{DiMari2025}) appears to offer a better fit for the Varese cohort than introducing a more flexible functional form for the covariates in the mixing proportion of the model proposed by Lambert et al. (2007) \cite{lambert2007}. Nonetheless, a more comprehensive comparison through additional simulation and experimental studies remains necessary. Finally, it is important to emphasize that the adequacy of cure models strongly depends on the data at hand, and no single model should be expected to perform best in all settings. Exploring different cure models provides valuable tools for practitioners working within the RS framework, which can be applied according to the cancer type under study. 

We found at least two reasons why the GAM model might have performed better than the Neural Network models. First, while the GAM model is penalized through the integral of the second derivative of the predictor \cite{wood2017}, the Neural Network model has no penalty for excessive wiggliness, leading to a higher risk of overfitting (cf. the out-of-sample log-likelihood in Table \ref{model_selection}). The second reason could be related to sample size. It is well known in the literature that training a Neural Network typically requires large sample sizes, and the larger the network, the more observations are required. Moreover, even when the sample size is large, although a general approximation theorem exists for Neural Networks, this result does not state how many hidden neurons are required to capture the exact form of the function we aim to estimate; in fact, wider is not necessarily better, and larger networks can be inefficient, hard to train, and poor at generalization. Using multiple layers allows us to represent functions in a more efficient and structured manner, avoiding the need for a very large number of neurons. A more thorough model selection strategy for the number of hidden neurons and layers could therefore lead to improved performance, and the exploration of alternative Neural Network architectures could be pursued in future work using larger datasets. 

Nevertheless, both the simulation and experimental studies suggest that, when only a few covariates are available (e.g. \textit{age at diagnosis}, \textit{period of diagnosis}, and \textit{sex}), which is often the case in population-based cancer studies, it is worthwhile to consider both the proposed Neural Network models and the GAM model instead of the standard GLM model. We found that, up to approximately $10{,}000$ observations, using four hidden neurons in the Neural Network model is sufficient to achieve better results than the standard GLM model. For larger sample sizes, increasing the size of the hidden layer may yield further improvements, although this requires additional investigation. Moreover, when the sample size is large and the number of covariates increases, we still recommend considering the proposed Neural Network model, as the GAM framework requires the user to explicitly specify interaction terms, which can become cumbersome and computationally demanding.

We derived important epidemiological indicators, which were useful for interpreting survival trends by age and year of diagnosis (cf. Table \ref{model_results} and Table \ref{time_to_cure}). The highest proportion of cured patients was registered in the periods 2003-2007 and 2008-2013, very likely due to the introduction of national screening programs starting from 2003 \cite{masseria2009}, which affected individuals between the ages of $45$ and $75$. The elderly were the ones having the lowest cure proportions, being the ones less prone to heal from chronic diseases due to biological reasons. Improvements in prognosis over time were evident across all age groups in the median survival time of fatal cases. In particular, the youngest individuals had longer survival times compared to those in other age classes. As individuals aged, the prognosis for fatal cases worsened; however, it still improved over the diagnosis periods, consistent with advancement in the field of oncological medicine. These findings highlighted the overall positive trend in cancer survival outcomes, with notable variations observed across different age groups. 

Future research could focus, on one hand, on performing sensitivity analyses on the choice of the Weibull distribution and, on the other hand, on extending the model-based approach considered in Di Mari et al. (2025) \cite{DiMari2025} to incorporate the flexible covariate functions used in this work. Moreover, while we proposed to use Neural Networks and GAMs to model the cure fraction, other models within the field of machine learning known to be very flexible could also be applied. Exploring methods such as Random Forests \cite{breiman2001}, XGBoost \cite{chen2016}, and Support Vector Machines \cite{cortes1995} may yield further improvements.

% \section{Competing interests}
% No competing interest is declared.

% \section{Author contributions statement}

% F.D.M., R.R., and R.D.A. designed the methodology; S.R. and G.T. provided and pre-processed the data; F.D.M. conducted the simulation and experimental studies and wrote the manuscript; all authors reviewed the manuscript.

\begin{acknowledgement}
The simulation study was performed thanks to the supercomputer TeraStat2 (\hyperlink{https://www.dss.uniroma1.it/it/HPCTerastat2}{https://www.dss.uniroma1.it/it/HPCTerastat2}).
\end{acknowledgement}

\bibliographystyle{vancouver}
\bibliography{references}
\end{document}